\documentclass[aps,pra,a4paper,amssymb,amsmath,superscriptaddress,twocolumn,showpacs]{revtex4-1}

\usepackage{amsmath, amssymb, amsfonts, amsthm}
\usepackage[english]{babel}
\usepackage{bm}
\usepackage{mathrsfs}
\usepackage{enumerate}
\newcommand{\bdm}{\begin{displaymath}}
\newcommand{\edm}{\end{displaymath}}
\newcommand{\bdn}{\begin{eqnarray}}
\newcommand{\edn}{\end{eqnarray}}
\newcommand{\bay}{\begin{array}{c}}
\newcommand{\eay}{\end{array}}
\newcommand{\ben}{\begin{enumerate}}
\newcommand{\een}{\end{enumerate}}
\newcommand{\beq}{\begin{equation}}
\newcommand{\eeq}{\end{equation}}
\newcommand{\bml}[1]{\begin{multline} #1 \end{multline}}
\newcommand{\bmln}[1]{\begin{multline*} #1 \end{multline*}}
\newcommand{\f}{\frac}

\newcommand{\R}{\mathbb{R}}

\newcommand{\ci}{\mathbb{C}}

\newcommand{\lf}{\left}
\newcommand{\ri}{\right}

\newcommand{\av}{\mathbf{a}}

\newcommand{\xv}{\mathbf{x}}
\newcommand{\Kv}{\mathbf{K}}

\newcommand{\yv}{\mathbf{y}}

\newcommand{\yyv}{\mathbf{Y}}

\newcommand{\qv}{\mathbf{q}}

\newcommand{\rv}{\mathbf{r}}
\newcommand{\sv}{\mathbf{s}}
\newcommand{\tthv}{\mathbf{\Theta}}
\newcommand{\omv}{\bm{\omega}}

\newcommand{\tv}{\mathbf{t}}

\newcommand{\kv}{\mathbf{k}}

\newcommand{\kkv}{\breve{\mathbf{K}}}

\newcommand{\pv}{\mathbf{p}}

\newcommand{\diff}{\mathrm{d}}

\newcommand{\ve}{\varepsilon}

\newcommand{\tx}{\textstyle}
\newcommand{\disp}{\displaystyle}

\newtheorem*{teo}{Theorem}%[section]

\newenvironment{rem}{\vspace{0,1cm} \noindent \textit{Remark}}

\newcommand{\be}{\begin{equation}}
\newcommand{\ee}{\end{equation}}
\newcommand{\bey}{\begin{eqnarray}}
\newcommand{\eey}{\end{eqnarray}}

\newcommand{\ba}{\begin{eqnarray}}
\newcommand{\ea}{\end{eqnarray}}

\newcommand{\donothing}[1]{}
\newcommand{\n}{\noindent}
\newcommand{\vs}{\vspace{0.5 cm}}

\usepackage{bbm}

\begin{document}

%\LARGE
%\textsc{Notes}
%\normalsize

%\medskip

\title{Energy lower bound for the unitary $N+1$ fermionic model}

\author{M. Correggi}
\affiliation{Dipartimento di Matematica e Fisica, Universit\`a degli Studi Roma Tre, L.go San Leonardo Murialdo 1, 00146 Roma, Italy}
\author{D. Finco}
\affiliation{Facolt\`a di Ingegneria, Universit\`a Telematica Internazionale Uninettuno, Corso V. Emanuele II 39,  00186 Roma, Italy}
\author{A. Teta}
\affiliation{Dipartimento di Matematica, ``Sapienza'' Universit\`a di Roma, P.le A. Moro 5, 00185 Roma, Italy}

\date{\today}


\begin{abstract}
We consider the stability problem for a unitary $N+1$ fermionic model, i.e., a system of $N$ identical fermions interacting via zero-range interactions with a different particle, in the case of infinite two-body scattering length.  We present a slightly more direct and simplified proof of a recent result obtained in \cite{CDFMT}, where a sufficient stability condition is proved under a suitable assumption on the mass ratio. 
\end{abstract}	

% \date{June 10th, 2011}

\maketitle

%\newpage

%\vs

\section{Introduction and Main Result}
The  study of the quantum mechanical many-body problem with pairwise zero-range interactions has received a considerable attention in recent years as an effective model describing the behavior of cold atoms near the BEC/BCS crossover (\cite{bh}, \cite{cw} and references therein). %This is essentially due to the recently achieved  possibility to realize experimental conditions where the interaction is well described by a zero-range force, in particular in the unitary limit. Roughly speaking,  unitary limit means  that the two-body interaction is characterized by a zero-energy resonance or, equivalently, by an infinite value of the scattering length.    
The correct definition of the model, the occurrence of the Efimov effect and the analysis of the stability problem, i.e., the existence of a lower bound for the Hamiltonian, have been widely studied both in the physical (\cite{cmp}, \cite{ct}, \cite{km}, \cite{ms}, \cite{tc}, \cite{wc1}, \cite{wc2}) and in the mathematical (see, e.g., \cite{CDFMT}, \cite{DFT}, \cite{FT}, \cite{m}) literature. 
It is well known that in the two-body case the entire class of Hamiltonians with zero-range interaction can be constructed  and the spectral properties are completely characterized (\cite{al}) while, on the opposite, for more than two particles an explicit characterization is still lacking.  
Proceeding by analogy with the two-body case, one can construct a physically reasonable class of Hamiltonians usually called Skornyakov Ter-Martirosyan (STM) operators. As a matter of fact, such operators are symmetric but (in general) are neither self-adjoint nor bounded from below. This happens, for instance, in the case of three identical  bosons, where it was shown in \cite{fm} that  the STM operator admits self-adjoint extensions which can be explicitly constructed but they are all unbounded from below and therefore the system is unstable. 

Here we are interested in the stability problem in the fermionic case, that is when fermions of different species interact among themselves. For the most  general system composed by a mixture of $N$ identical fermions of one type (with mass $m_1$) and $K$ identical fermions of a different type (with mass $m_0$),  the stability problem for the corresponding STM operator is open and some results are available only in special cases. For instance, a system composed by two identical fermions plus a different particle  is known to be stable if (and only if) the mass ratio 
\be\label{massra}
\alpha = \f{m_1}{m_0}
\ee
 is smaller than the critical value $13.607$ (see, e.g., \cite{bh}, \cite{CDFMT}. Further results are available in the $3+1$ case only (\cite{cmp}).

%\n
In this note we review a  result obtained in \cite{CDFMT} on the stability of a system composed by $N$, with $N \geq 2$,  identical fermions plus a different particle. In particular it is shown that  stability occurs if  
\be\label{stacond}
\alpha \leq \alpha_c (N)
\ee
where $\alpha_c (N)$ is the solution of the following equation
\be\label{lam}
\Lambda(\alpha,N) := \tx\f{2}{\pi}(N\!-\!1) \left(\!\f{1+\alpha}{\alpha} \!\right)^{\!2} \!\left[\! \f{\alpha}{\sqrt{1+2\alpha}}- \arcsin \left(\!\f{\alpha}{1+\alpha}\!\right)\right] =1
\ee
Notice that for each $N$ the function $\Lambda(\alpha,N)$ is increasing,  goes to infinity for $\alpha \rightarrow \infty$ and $\Lambda(0,N)=0$. So there is exactly one solution $\alpha_c(N)>0$ of  (\ref{lam}) and moreover $\Lambda(\alpha,N) <1$ for $\alpha < \alpha_c(N)$. We remark that only for $N=2$ the condition (\ref{stacond}) is optimal, i.e., $\alpha_c (2) = 13.607$, and therefore the result provides a rigorous proof of what is already known in the physical literature. 

For $N>2$ the condition is surely not optimal since, as it will be clear from the proof, the role of the antisymmetry is only partially exploited. Nevertheless we believe that the result can be of some interest since (\ref{stacond}) gives a first sufficient stability condition which, apparently, was not known before. Some numerical values of $\alpha_c(N)$ are listed here
\begin{eqnarray}
\alpha_c(3) &=& 5.291,	\nonumber	\\
\alpha_c(8)&=& 1.056, 	\nonumber	\\
\alpha_c(9)&=& 0.823, \ldots
\end{eqnarray}
In particular this means that, in the case of equal masses, the system is stable if $N\leq8$.

%\n
Our aim here is to present a slightly more direct and simplified proof of the above stability condition, limiting ourselves to the simpler but relevant unitary case (see below). We hope that such a presentation could help in clarifying the points of the proof which must be improved to obtain a more satisfactory stability condition.

%\n
Let us consider the formal Hamiltonian for a system of $N$ identical fermions with mass $m_1$ and a different particle with mass $m_0$ (we set $\hbar=1$)
\be\label{formal}
\tilde{H}= -\tx\f{1}{2 m_1} \sum_{i=1}^N \Delta_{\xv_i} - \f{1}{2 m_0} \Delta_{\xv_0} + \mu\sum_{i=1}^N \delta(\xv_0 - \xv_i)\,.
\ee
The parameter $\mu \in \R$ is a coupling constant which must be properly renormalized in order to give a precise meaning to the expression (\ref{formal}). We introduce center of mass and relative coordinates 
\be
\xv_c= \tx
\f{1}{M} \left( m_0 \xv_0 + m_1 \sum_{i=1}^N \xv_i \right) , \;\;\;\;\;\; \yv_i=\xv_0 - \xv_i
\ee
where $M=m_0 + N m_1$.  %, with inverse 
%\be\label{invtr}
%\xv_0=\tx  \xv_c+ \f{m_1}{M} \sum_{k=1}^N   \yv_k\,, \;\; \xv_i = \xv_c +\left( \f{m_1}{M} -1\!\right)\! \yv_i + \f{m_1}{M} \sum_{k\neq i}^N \yv_k
%\ee
One has
\bml{
\tilde{H}=- \tx\f{1}{2 M} \Delta_{\xv_c} - \f{m_0 +m_1}{2 m_0 m_1} \sum_{i=1}^N \Delta_{\yv_i} \\
 - \tx\f{1}{m_0} \sum_{i<j}\nabla_{\yv_i} \cdot \nabla_{\yv_j}  +\mu \sum_{i=1}^N \delta(\yv_i)\,.
}
Therefore in the center of mass reference frame the system  is described by the formal Hamiltonian 
\be\label{formal2}
\overline{H}= H_0 + \bar{\mu} \tx\sum_{i=1}^N \delta(\yv_i)
\ee
where $\;\bar{\mu}= \f{2 m_0 m_1}{m_0 + m_1} \mu\;$ and $H_0$ is the free Hamiltonian
\be
H_0= - \tx\sum_{i=1}^N \Delta_{\yv_i}  - \f{2 \alpha}{1+\alpha} \sum_{i<j}\nabla_{\yv_i} \cdot \nabla_{\yv_j} %\, \;\;\;\;\;\;\;\; \alpha= \f{m_1}{m_0}
\ee
with $\alpha$ the mass ratio (\ref{massra}). The formal Hamiltonian (\ref{formal2}) can be given a precise meaning as a, possibly self-adjoint,  operator in the natural Hilbert space $L^2_{\mathrm{ant}}(\R^{3N})$, i.e.,  the Hilbert space of antisymmetric, square integrable functions on $\R^{3N}$. More precisely, it is by definition a non trivial (self-adjoint) extension of the free Hamiltonian $H_0$ restricted to smooth functions vanishing on the set
\be\label{omega}
\Omega\; = \tx\bigcup_{i \in \{1,\ldots,N\} } \left\{ \yyv \equiv(\yv_1, \ldots, \yv_N) \in \R^{3N} \,|\, \yv_i =0    \right\}.
\ee
As we already mentioned, among all possible extensions a special role is played by the STM operator, due to the fact that it is the natural generalization of the well known  Hamiltonian with zero-range interaction in the two-body case. 

For two (different) particles, extracting the center of mass motion and denoting by $\xv$ the relative coordinate, the domain of the operator consists of functions $\psi \in L^2(\R^3)$, which are regular for $\xv \neq 0$ and satisfy the following boundary condition as $ |\xv|\rightarrow 0
 $
\be\label{bc1d}
\psi(\xv) = \left(\f{1}{|\xv|} - \f{1}{a} \right) q +o(1),
\ee
where $q \in \ci$ depends on $\psi$ and $a \in \R$ has the physical meaning of  a scattering length.

\n
Moreover, the Hamiltonian acts as the free Hamiltonian for $|\xv|\neq 0$.

%\n
The STM extension $H_a$ in our fermionic $N+1$-particle system is defined in an analogous way. In the original coordinate system $\xv_0, \xv_1, \ldots ,\xv_N$, and extracting the center of mass motion,  the domain $D(H_a)$ is made of functions $\phi$ defined on the set
\be
\{(\xv_0,  \ldots ,\xv_N) \in \R^{3(N+1)} \,|\, \xv_c =0\}
\ee
antisymmetric under the exchange of any pair of fermions, regular for $\xv_0 \neq \xv_i$, $i=1,\ldots,N$. The standard formulation of the  boundary condition satisfied as $|\xv_0 - \xv_i|\rightarrow 0$ is (see, e.g., \cite{wc2})
\bml{\label{bcx}
\phi(\xv_0,\ldots ,\xv_N) =\! \left(\! \f{1}{|\xv_0 \!-\! \xv_i|} - \f{1}{a} \right)\! (-1)^{i+1} Q(\rv_{0i}, \breve{\xv}_i) \\
+ o(1)
}
where $Q$ is a given function depending on $\phi$ and
\ba
&&\rv_{0i}=\f{m_0 \xv_0 + m_1 \xv_i}{m_0 + m_1}\\
&&\breve{\xv}_i= (\xv_1,\ldots,\xv_{i-1},\xv_{i+1},\ldots,\xv_N)
\ea
Notice that in the limiting procedure defining the boundary condition \eqref{bcx} the vectors $\rv_{0i}$ and $\breve{\xv}_i$ are kept fixed. 
Passing to the relative coordinates $\yyv=(\yv_1,\ldots ,\yv_N)$, the wave function $\psi(\yyv)=\phi(\xv_0(\yyv),\ldots ,\xv_N(\yyv))$ is an element of $ L^2_{\mathrm{ant}}(\R^{3N})$, regular for $\yyv \notin \Omega$. Moreover, setting $\hat{Q}(\yyv) = Q(\rv_{0i}(\yyv),\breve{\xv}_i (\yyv))$, the boundary condition \eqref{bcx} satisfied as $|\yv_i|\rightarrow 0$ now reads
\bml{ \label{boucon}
\psi(\yyv)=\left( \f{1}{|\yv_i|} - \f{1}{a} \right) (-1)^{i+1} \hat{Q}(\yyv) + o(1) \\
= \left( \f{1}{|\yv_i|} - \f{1}{a} \right) (-1)^{i+1} \hat{Q}(\yyv)\big|_{\yv_i=0} \\
+(-1)^{i+1} \nabla_{\yv_i} \hat{Q}(\yyv) \big|_{\yv_i =0} \cdot \f{\yv_i}{|\yv_i|} +o(1)\\
\equiv \left( \f{1}{|\yv_i|} - \f{1}{a} \right) (-1)^{i+1}  \xi (\breve{\yv_i}) + 
\tthv_i (\breve{\yv}_i) \cdot \omv_i
+ o(1)
}
where $\xi \,:\, \R^{3N-3} \rightarrow \ci$ is an antisymmetric function, %depending on $\psi$, 
$\omv_i= \yv_i/|\yv_i|$ and $a \in \R$ is the  two-body scattering length corresponding to the interaction of a fermion with the different particle. 

\n
We remark that in \eqref{boucon} (i.e., the boundary condition written in the relative coordinates) an extra term appears, depending on the unit vector $\omv_i$, namely the direction along which the limit $ \yv_i \to 0 $ is taken. As a matter of fact, such a term does not contribute to the energy of the system (see Section II), yielding the same expression one would get if that term was absent (see also the Remark at the end of Section II).

\n
Furthermore,  $H_a$ acts as the free Hamiltonian outside the set $\Omega$, i.e.,
\be\label{azh}
(H_a \psi)(\yyv) = (H_0 \psi)(\yyv),	\qquad		\mbox{if } \yyv \in \R^{3N} \setminus \Omega.
\ee
 A special role is played by the parameter-free case of infinite scattering length, known as the unitary case. We shall denote by $H$ the corresponding STM extension, i.e., $ H := H_{\infty} $.

%\n
The main result discussed in this note is the following

\begin{teo}
\label{teo}
 In the unitary case the  energy form, i.e., the expectation value of the energy, is positive for $  \alpha\leq\alpha_c(N) $. More precisely,  for any $ \psi \in D(H) $,
\be\label{main}
 (\psi, H \psi) \geq 0,	\qquad \mbox{if}\;\;\; \alpha\leq\alpha_c(N).
\ee
This in particular implies stability for the unitary $N+1$ fermionic model.
\end{teo}

\n
In the next Section we derive a suitable expression for  the energy form. In Section III we start from such expression to explain the steps required to prove our result. In Section IV we briefly summarize the content of the paper. In Appendix A we collect some technical results useful to reformulate the domain and the boundary condition characterizing the Hamiltonian.

\section{Derivation of the energy form}

Here we derive a suitable expression for  the energy form. The key point is to represent the domain $ D(H) $  as the set of wave functions decomposing as 
\be\label{deco}
\psi = w + \mathcal G \xi 
\ee
where $ w $ is a smooth function and $ \mathcal G \xi $ contains the singular behavior prescribed in \eqref{boucon}. More precisely, $ \mathcal G \xi $ is the ``potential''  produced by the ``charge'' $\xi$ distributed on the planes $\{\yv_i =0\}$, i.e., 
\be\label{poten}
({\cal G} \xi)(\yyv)= \sqrt{\f{2}{\pi}} \, \sum_{j=1}^N \f{(-1)^{j+1}}{(2\pi)^{\f{3}{2}N} }
 \!\int \!\! \diff \Kv \, e^{i \Kv \cdot \yyv} G(\Kv) \,\hat{\xi}(\breve{\kv}_j),
\ee
with $ \Kv : = (\kv_1, \ldots, \kv_N) $, $\hat{f}$ denoting the Fourier transform of $f$ and 
\be\label{green}
G(\Kv)=\f{1}{\sum_i \kv_i^2 + \f{2 \alpha}{1+\alpha} \sum_{i<j} \kv_i \cdot \kv_j}.
\ee
Indeed, it is straightforward to verify that 
\ba\label{asy}
&&({\cal G} \xi)(\yyv)=\f{(-1)^{i+1} \xi(\breve{\yv}_i)}{|\yv_i|} - \big(\Gamma \xi \big) (\breve{\yv}_i) + o(1),
\ea
for $ |\yv_i| \rightarrow 0 $, where $\,\Gamma \xi\,$ is the inverse Fourier transform of 
\ba\label{Gam}
 \lf(\widehat{\Gamma \xi} \ri) (\breve{\kv}_i) &=& \tx\f{\sqrt{2\alpha +1}}{1+\alpha} (-1)^{i+1}L(\breve{\kv}_i) \hat{\xi}(\breve{\kv}_i) \nonumber\\
& - & \f{1}{2\pi^2} \disp\sum_{j, j\neq i} (-1)^{j+1} \int \!\! \diff\kv_i\, G(\Kv) \, \hat{\xi}(\breve{\kv}_j)
\ea
and
\be\label{elle}
L(\kv_1,\ldots,\kv_{N-1})= \sqrt{ \tx\sum_i \kv_i^2 + \tx\f{2\alpha}{1+2 \alpha} \sum_{i<j}\!\! \kv_i \cdot \kv_j         }\,.
\ee
Moreover, it is useful to note that the potential ${\cal G} \xi$ satisfy the equation
\be\label{distri}
\big( H_0 \,{\cal G} \xi \big)(\yyv)  = 4 \pi \sum_i (-1)^{i+1}\xi(\breve{\yv}_i) \, \delta(\yv_i)
\ee
in distributional sense and then, in particular,
\be\label{distri2}
\big( H_0 \,{\cal G} \xi \big)(\yyv)  =0,	\qquad		\mbox{if } \yyv \in \R^{3N} \setminus \Omega.
\ee
 The proof of \eqref{asy} and  \eqref{distri}  is postponed in Appendix \ref{bc}.
 
 \n
Using the decomposition \eqref{deco} and the asymptotic behavior \eqref{asy}, the boundary condition (\ref{boucon}) in the unitary case can be equivalently written as
\be\label{bc2}
\lim_{|\yv_i|\rightarrow 0} w (\yyv) = (\Gamma \xi)(\breve{\yv}_i) + \tthv_i (\breve{\yv}_i) \cdot \omv_i.
%\big( \widehat{w|_{\yv_i =0}} \big) (\breve{\kv}_i)= \big( \widehat{ \Gamma \xi }\big) (\breve{\kv}_i) + ..................................
\ee

We can now  derive the expression for the energy form. Taking into account \eqref{azh}, the decomposition \eqref{deco} and equation \eqref{distri2} we have 
\ba \label{for1}
&&(\psi,H\psi)= \lim_{\ve \rightarrow 0} \int_{|\yv_1|>\ve}\!\!\!\!\! \diff \yv_1 \ldots \!\!\int_{|\yv_N|>\ve}\!\!\!\!\! \diff \yv_N\, \bar{\psi} (\yyv) \big( H_0 \psi)(\yyv) \nonumber\\
&&=\lim_{\ve \rightarrow 0} \int_{|\yv_1|>\ve}\!\!\!\!\! \diff \yv_1 \ldots \!\!\int_{|\yv_N|>\ve}\!\!\!\!\! \diff \yv_N\, \big( \bar{w} + \overline{\cal G \xi} \big) (\yyv) \big( H_0 w)(\yyv)
\nonumber\\
&&= (w,H_0 w)\nonumber\\
&&+ \lim_{\ve \rightarrow 0} \int_{|\yv_1|>\ve}\!\!\!\!\! \diff \yv_1 \ldots \!\!\int_{|\yv_N|>\ve}\!\!\!\!\! \diff \yv_N\,  \big(\overline{\cal G \xi} \big) (\yyv) \big( H_0 w)(\yyv).
\ea
In the last integral of \eqref{for1} we apply Green's identities. Denoting $S^2_i=\{\yv_i \in \R^3 \,|\, |\yv_i|=1\}$, for $\ve \rightarrow 0$ we have 
\ba\label{for2}
&& \int_{|\yv_1|>\ve}\!\!\!\!\! \diff \yv_1 \ldots \!\!\int_{|\yv_N|>\ve}\!\!\!\!\! \diff \yv_N\,  \big(\overline{\cal G \xi} \big) (\yyv) \big( H_0 w)(\yyv)\nonumber\\
&&=-\ve^2 \sum_{i=1}^N \int_{\R^{3(N-1)} }\!\!\!\! \!\diff \breve{\yv}_i \!\!\int_{S^2_i}\!\!\! \diff \sigma(\omv_i) \f{\partial \big(\overline{\cal G \xi} \big)}{\partial |\yv_i|}  \bigg|_{|\yv_i|=\ve} \!\! w\big|_{|\yv_i|=\ve} \nonumber\\
&& \;\;\;\;\;\;\;\; \;\;\;\;\;\;\;\; \;\;\;\;\;\;\;\; \;\;\;\;\;\;\;\; \;\;\;\;\;\;\;\; \;\;\;\; \;\;\;\;\;\;\;\; \;\;\;\;\;\;\;\; \;\;\;\;\;\;\;\; \;\;\;\;\;\;\;\;+ \;o(1)\nonumber\\
&&=\sum_{i=1}^N  (-1)^{i+1} \!\!\int_{\R^{3(N-1)} }\!\!\!\! \!\diff \breve{\yv}_i \, \overline{\xi}(\breve{\yv}_i) \!\! \int_{S^2_i}\!\!\! \diff \sigma(\omv_i) \,w\big|_{|\yv_i|=\ve}  \!\!\! +o(1) \nonumber\\
&&=\sum_{i=1}^N  (-1)^{i+1} \!\!\int_{\R^{3(N-1)} }\!\!\!\! \!\diff \breve{\yv}_i \, \overline{\xi}(\breve{\yv}_i)\int_{S^2_i}\!\!\! \diff \sigma(\omv_i) \, \bigg(  (\Gamma \xi)(\breve{\yv}_i) \nonumber\\
&& \;\;\;\;\;   \;\;\;\;\;\;\;\; \;\;\;\;\;\;\;\; \;\;\;\;\;\;\;\; \;\;\;\;\;\;\;\; \;\;\;\;  \;\;\; +  \;  \tthv_i (\breve{\yv}_i) \cdot \omv_i  \bigg)   +o(1),
\ea
where we have used equation \eqref{distri2}, the asymptotics \eqref{asy} and the boundary condition \eqref{bc2}.  Observe that in \eqref{for2} the term $o(1)$ contains surface integrals going to zero for $\ve \rightarrow 0$. Note also that, due to the integration over the unit sphere $S^2_i$, the contribution  of the term $\tthv_i (\breve{\yv}_i) \cdot \omv_i $  vanishes. Taking into account  \eqref{for1} and  \eqref{for2}, we find
\ba\label{for3}
\!\!\!\!\!\!\!\!(\psi, H \psi)&=& (w, H_0 w)  \nonumber\\
&+& \;4 \pi \sum_{i=1}^N  (-1)^{i+1} \!\!\int_{\R^{3(N-1)} }\!\!\!\! \!\diff \breve{\kv}_i \, \overline{\hat{\xi}}(\breve{\kv}_i) \, \widehat{(\Gamma \xi)} (\breve{\kv}_i).
\ea

\n
Inserting in \eqref{for3} the explicit expression \eqref{Gam} of $\widehat{\Gamma \xi}$ and  
exploiting the antisymmetry property of the charge $\xi$ we finally obtain 
\be\label{forma}
(\psi, H\psi) = (w, H_0 w) + \f{2N}{\pi} \,\Phi (\xi),
\ee
where the quadratic form $\Phi$ is defined by
\beq\label{forxi}
	\Phi(\xi)= \int\!\! \diff\kkv\, \lf( \Phi_1(\xi;\kkv) + (N-1)\Phi_2(\xi;\kkv)  \ri),    
\eeq
%d\kv_1\ldots d\kv_{N-2} \Big(  \Phi_1(\xi;\kv_1,\ldots,\kv_{N-2})+ \Phi_2(\xi;\kv_1,\ldots,\kv_{N-2}) \Big) \\ %2 \pi^2 \f{\sqrt{2+\alpha}}{1+\alpha} \int\!\! d\sv \, L(\sv, \kv_1,\ldots,\kv_{N-2}) |\hat{\xi}(\sv,\kv_1,\ldots,\kv_{N-2})|^2 \nonumber\\
%&&+(N-1)\! \int\!\! d\sv d\tv \, G(\sv,\tv, \kv_1, \ldots,\kv_{N-2}) \, \overline{\hat{\xi}}(\sv, \kv_1, \ldots,\kv_{N-2}) \hat{\xi}(\tv, \kv_1, \ldots,\kv_{N-2}) \Bigg)\\
 %$ \kkv=(\kv_1,\ldots,\kv_{N-2})$ 
 with
 \bdm
 	\kkv=(\kv_1,\ldots,\kv_{N-2}),
\edm
\ba
\Phi_1(\xi;\kkv) &=& 2 \pi^2 \tx\f{\sqrt{1+2\alpha}}{1+\alpha} \disp\int\!\! \diff\sv \, L(\sv, \kkv) |\hat{\xi}(\sv,\kkv)|^2\,, \label{fi1}\\
\Phi_2(\xi;\kkv) &=&\! \int\!\! \diff\sv \diff\tv \, G(\sv,\tv, \kkv) \, \overline{\hat{\xi}}(\sv, \kkv) \hat{\xi}(\tv, \kkv)\,, \label{fi2} 
\ea

\n
Note that in the special case $N=2$ the extra variables $\breve{\bf K}$ are absent.
\vs

\begin{rem}
\mbox{}\\
As we have seen in the above computation, the presence of the term $\tthv_i (\breve{\yv}_i) \cdot \omv_i $ in \eqref{boucon} is irrelevant, in the sense that it does not contribute to the energy and therefore it can be dropped from the outset. In our opinion, this simply means that in the boundary condition the limit for $|\xv_0 - \xv_i|\rightarrow 0$ does not depend on the direction along which the limit is taken.  % must be understood as independent of the direction $\omv_i$.  
In order to stress this fact, \eqref{bcx} should be more correctly written in the following way 
\bml{\label{bcxx}
\f{1}{4\pi} \int_{S^2_i} \!\!\!\! \diff \sigma(\omv_i) \, \phi(\xv_0,\ldots ,\xv_N)  
\\ = \! \left(\! \f{1}{|\xv_0 \!-\! \xv_i|} - \f{1}{a} \right)\! (-1)^{i+1} \, Q(\rv_{0i}, \breve{\xv}_i) + o(1)
}
as $|\xv_0 - \xv_i|\rightarrow 0$. 
\end{rem}

\section{Positivity of the Energy}

From the above expression (\ref{forma}) we see that the Theorem is proved  if we can show positivity of $\Phi$. Since the term $\Phi_1$ is positive, the problem is reduced to show that 
$$ (N-1) \Phi_2 \geq  - c\, \Phi_1 $$ 
for some constant $c \leq 1$.   A proof of this fact will be given here and for the sake of clarity it will be divided in several, but elementary, steps. The strategy will be the reduction of the form $ \Phi_2 $ to one which can be diagonalized. This first requires a suitable change of variables; then we exploit the rotational symmetry of $ \Phi_2 $ to perform a partial wave decomposition; once the additional degrees of freedom are dropped, the problem reduces to bound from below a two-particle energy, which can be diagonalized by means of the Fourier transform; to conclude the proof it suffices then to go back to the original expression and show that, if the condition $ \alpha \leq \alpha_c $ is satisfied, $ \Phi $ is positive. It is worth stressing that at the last stage of the proof (see, e.g., \eqref{st1}) the fermionic symmetry of the charges $ \xi $ is totally neglected, in order to diagonalize the expression. This is clearly not optimal and an improvement of the condition \eqref{stacond} would require a different approach. In fact the change of variables itself (see \eqref{changevariable}), which is the starting point of our analysis, make the antisymmetric requirement not apparent and therefore should probably be avoided if one wants to track down the role of the fermionic antisymmetry.

\subsection{Change of variables} 

\n
We define
\be
\pv=\sv+ \f{\alpha}{2 +\alpha} \sum_{i=1}^{N-2} \kv_i \,, \qquad \qv=\tv+\f{\alpha}{2+\alpha} \sum_{i=1}^{N-2} \kv_i \,,
\ee
and therefore we obtain
\ba\label{fi11}
\Phi_1(\xi;\kkv) &=& 2 \pi^2\!\! \int\!\! \diff \pv \, \sqrt{ \tx\f{1+2\alpha}{(1+\alpha)^2} \pv^2 + D(\kkv)} \,\, |\eta(\pv,\kkv)|^2\,, \nonumber \\
\Phi_2(\xi;\kkv) &=& \int\!\! \diff \pv \diff \qv \, \f{\overline{\eta} (\pv,\kkv) \eta(\qv,\kkv)}{\pv^2 +\qv^2 +\f{2\alpha}{1+\alpha}\, \pv \cdot \qv +D(\kkv)}\,, \label{fi22}
\ea
where
\be
\label{changevariable}
\eta (\pv,\kkv)= \hat{\xi}\left(\! \pv - \tx\f{\alpha}{1+2\alpha} \sum \kv_i, \kkv \!\right)\,,
\ee
\bdm
D(\kkv)=\tx\frac{1+3\alpha}{(1+\alpha)(1+2 \alpha)} \lf(  \sum k_i^2 + \f{2 \alpha}{1+3\alpha} \tx\sum_{i < j} \kv_i \cdot \kv_j \ri)\,.
\edm

%In the next steps we show how to obtain the estimate from below of (\ref{fi22}) in terms of (\ref{fi11}). 

\subsection{Expansion in spherical harmonics}

%\n
For any $f\in L^2(\R^3)$ we consider the  expansion
\beq\label{exsh}
f(\pv) = \sum_{l=0}^{\infty}\sum_{m=-l}^l f_{lm}(p) Y_{l}^m(\theta_p, \phi_p)\,
\eeq
where $\pv=(p, \theta_p,\phi_p)$  and $Y_{l}^m$ denotes the spherical harmonics of order $l,m$ with $ l =0,1,\ldots $ and $ m = -l, \ldots, l $. Moreover, we denote by $P_l$ the Legendre polynomial of order $l$ explicitly given by
\beq\label{leg}
P_l(y) = \f{1}{2^l l!} \f{d^l}{dy^l} (y^2 -1)^l \,,	\qquad y \in [-1,1]\,.
\eeq
Using the above expansion we derive the following decomposition of $\Phi_2$ in each subspace of fixed angular momentum $l$:
\ba
\Phi_2(\xi;\kkv)&=&\! 2\pi  \sum_{l=0}^{\infty} \!\sum_{m=-l}^l 
\!\int_0^{\infty} \!\!\!\! \!\diff p\!\!\int_0^{\infty}\!\!\!\!\! \diff q\, \: \overline{\eta_{lm}}(p,\kkv)  \eta_{lm}(q,\kkv) 	\nonumber	\\
				& \times & \!\int_{-1}^1 \!\!\!\! dy\, \frac{p^2 q^2\, P_l(y)}{p^2 + q^2 +\frac{2\alpha}{1+\alpha} p q y +D(\kkv)}\nonumber\\
&=:& \sum_{l=0}^{\infty}\sum_{m=-l}^l G_l(\eta_{lm};\kkv).
\ea

%\vs
%\n
%- Behavior for each fixed $l$.

%\n
It turns out that (for details see \cite[Lemma 3.2]{CDFMT})  
\ba\label{even}
G_l(\eta_{lm};\kkv)		 \geq 0\,, \qquad & &\mbox{for $l$ even}, \\
0 \geq G_l(\eta_{lm};\kkv)  \geq G_l^0(\eta_{lm})\,,	\qquad & &\mbox{for $l$ odd}\label{odd},
\ea
where $G^0_l$ is defined by
\bml{
 G_l^0(\eta_{lm})= 2\pi  \!\int_{-1}^1 \!\!\!\! \diff y\, P_l(y) \times 	\\
 \times \!\int_0^{\infty} \!\!\!\! \! \diff p\!\!\int_0^{\infty}\!\!\!\!\! \diff q\,\, \f{ p^2 \, \overline{\eta_{lm}}(p,\kkv)\, q^2 \,\eta_{lm}(q,\kkv) }{p^2 + q^2 +\frac{2\alpha}{1+\alpha} p q y}\,.
}
From (\ref{even}) and (\ref{odd}) we then get 
\be\label{st1}
\Phi_2(\xi;\kkv) \geq \sum_{l\, \mathrm{odd}}\sum_{m=-l}^l G_l^0(\eta_{lm}).
\ee

\subsection{Diagonalization}

%\n
Let us define
\beq \label{formula2}
 g^{\sharp} (k) = \frac{1}{\sqrt{2\pi}} \int \diff x \, e^{-ikx}\, e^{2x} \, g(e^x).
\eeq
Then
\bmln{
	G_l^0(g)=2\pi  \!\int_{-1}^1 \!\!\!\! \diff y\, P_l(y) \times 	\\
	 \times \int \!\! \diff x_1 \diff x_2\,\,\f{ e^{3 x_1}\overline{g} (e^{x_1})\, e^{3 x_2} g(e^{x_2})}
{ e^{2 x_1} + e^{2x_2} +\f{2\alpha }{1+\alpha} y \,e^{x_1 +x_2}  } \\
 = \; \pi  \!\int_{-1}^1 \!\!\!\! \diff y\, P_l(y)
\int  \diff x_1 \diff x_2\,\,\f{ e^{2 x_1}\overline{g} (e^{x_1})\, e^{2 x_2} g(e^{x_2})}{
 \cosh (x_1-x_2) +\f{\alpha }{1+\alpha}y  }\,.
}
The last integral is a convolution and therefore can be diagonalized  by means of Fourier transform. 
Using the explicit Fourier transform of the kernel (see, e.g., \cite{erdely}) we find for $l$ odd
\ba
&&G_l^0(\eta_{lm})= \int \!\! \diff k\, S_l (k) \,|\eta_{lm}^{\sharp}(k,\kkv)|^2\,,\\
&&S_l (k)= - \tx\f{\pi^2}{ \sinh \lf(\f{\pi}{2} k\ri)} 
\disp\int_{-1}^1 \!\!\! \diff y \, P_l(y) \, \tx\f{\sinh \lf( k \arcsin  \f{ \alpha }{1+\alpha} y\ri) }{ \cos \lf(  \arcsin  \f{ \alpha }{1+\alpha}y\ri) \, }
\ea
and the estimate (\ref{st1}) becomes
\be\label{st2}
\Phi_2(\xi;\kkv) \geq \sum_{l\, \mathrm{odd}}\sum_{m=-l}^l \int \!\! \diff k\, S_l (k) \,|\eta_{lm}^{\sharp}(k,\kkv)|^2\,.
\ee

%\vs
%\n
\subsection{Bound from below} 

%\n
We notice that, for any fixed $l$, $S_l(k)$ is an even, $C^{\infty}$ function of $k$ and $\lim_{k \rightarrow \infty}S_l(k)=0$. Furthermore 
for $l$ odd we can show that $S_l (k)$ is an  increasing function of $l$ for any fixed $k$ (for details see \cite[Lemma 3.5]{CDFMT}). Then
\be
S_l(k) \geq S_1(k)\,.
\ee
Moreover it is easy to see that 
\be
S_1 (k) \geq S_1(0) =4 \pi \tx\f{1+\alpha}{\alpha} \lf[ \f{\sqrt{1 + 2 \alpha}}{\alpha} \arcsin \left(\tx\f{\alpha}{1+\alpha} \right) - 1 \ri]
\ee
where $S_1(0) <0$.  Therefore from  (\ref{st2}) we have
\be\label{st3}
\Phi_2(\xi;\kkv) \geq - |S_1(0)| \sum_{l=0}^{\infty} \sum_{m=-l}^l \int \!\! \diff k\, |\eta_{lm}^{\sharp}(k,\kkv)|^2\,,
\ee
which can be rewritten in such a way to reconstruct the term $\Phi_1$. Indeed
\ba
&&\int \!\! \diff k\, |\eta_{lm}^{\sharp}(k,\kkv)|^2 = \int_{\R} \!\! \diff x\, e^{2x} |\eta_{lm}(e^x,\kkv)|^2 \nonumber	\\
&&\leq \tx\f{1+\alpha}{\sqrt{1 + 2 \alpha}} \disp\int_0^{\infty} \!\!\! \diff p\, p^2 \sqrt{ \tx\f{1+2\alpha}{(1+\alpha)^2} p^2 + D(\kkv)} \, |\eta_{lm}(p,\kkv)|^2\,. \nonumber
\ea
Using this estimate in (\ref{st3}) we find
\ba\label{st4}
&&\Phi_2(\xi;\kkv) \geq - |S_1(0)| \tx\f{1+\alpha}{\sqrt{1 + 2 \alpha}} \times \nonumber	\\
&& \times \disp\sum_{l=0}^{\infty} \sum_{m=-l}^l  \disp\int_0^{\infty} \!\!\! \diff p\, p^2 \sqrt{ \tx\f{1+2\alpha}{(1+\alpha)^2} p^2 + D(\kkv)} \, |\eta_{lm}(p,\kkv)|^2 \nonumber\\
&&= - |S_1(0)| \tx\f{1+\alpha}{\sqrt{1 + 2 \alpha}} \disp\int \!\! \diff \pv \, \sqrt{ \tx\f{1+2\alpha}{(1+\alpha)^2} p^2 + D(\kkv)} \, |\eta(\pv,\kkv)|^2 \nonumber\\
&&=  - |S_1(0)| \tx\f{1+\alpha}{2 \pi^2 \sqrt{1 + 2 \alpha}} \, \Phi_1(\xi;\kkv) \nonumber\\
&&= - (N-1)^{-1} \Lambda(\alpha,N) \, \Phi_1(\xi;\kkv)\,. \nonumber
\ea
We are now in position to conclude the proof of the Theorem. From (\ref{forma}), (\ref{forxi}) and the inequality above, we get
\bml{
(\psi, H\psi) \geq  \f{2 N}{\pi} \Phi (\xi) \\
\geq \f{2 N}{\pi} \lf(1 - \Lambda(\alpha,N) \ri) \disp\int \!\!\diff \kkv \,  \Phi_1(\xi;\kkv),
}
and taking $\alpha\leq \alpha_c (N)$ we obtain the desired result  $(\psi, H\psi) \geq 0$.

\section{Conclusions}

We have reported on a derivation of a sufficient condition on the mass ratio for the stability for the unitary $N+1$ fermionic model. Such a condition, which is optimal only in the two-particle case, is nevertheless non trivial for generic $N$. For istance it provides stability in the case of equal masses up to $N=8$. We have also described the main steps of the proof, enlightening the points  to be improved to get a more refined stability condition. 

\appendix

\section{Properties of the potential $\mathcal G \xi$ \label{bc}}

We first prove the asymptotic expression \eqref{asy} for the potential $\mathcal G \xi$ for $|\yv_i|\rightarrow 0$. 
It is convenient to isolate the $i$-th term of the sum in (\ref{poten})
\ba\label{A}
 A : &=& \sqrt{\f{2}{\pi}}\f{(-1)^{i+1}}{(2 \pi)^{\f{3}{2}N}} \int\!\!\diff\breve{\kv}_i \, e^{i \breve{\yv}_i \cdot \breve{\kv}_i} \hat{\xi} (\breve{\kv}_i) \int \!\!\diff \kv_i \, e^{i \yv_i \cdot \kv_i} G(\Kv)\nonumber\\
&=& \sqrt{\f{2}{\pi}}\f{(-1)^{i+1}}{(2 \pi)^{\f{3}{2}N}} \int\!\!\diff\breve{\kv}_i \, e^{i \breve{\yv}_i \cdot \breve{\kv}_i} \hat{\xi} (\breve{\kv}_i) \times	\nonumber	\\
&\times& \left( \int \!\!\diff \kv_i \, \f{e^{i \yv_i \cdot \kv_i}}{\kv_i^2} 
- \int \!\!\diff \kv_i \, \f{e^{i \yv_i \cdot \kv_i} (\kv_i \cdot \av +b)}{\kv_i^2 ( \kv_i^2 + \kv_i \cdot \av +b)}
\right)\!,
\ea
where 
\ba
&&  \av = \tx\f{2\alpha}{1 +\alpha} \sum_{l, l\neq i} \kv_l  \\
&& b= \breve{\kv}_i^2 +\tx\f{2\alpha}{1+\alpha} \sum_{j <l; j,l \neq i} \kv_j \cdot \kv_l .
\ea
By an explicit computation we find 
\ba\label{intg}
&&\int \!\!d \kv_i \, \f{e^{i \yv_i \cdot \kv_i}}{\kv_i^2} = \f{2\pi^2}{|\yv_i|}  
\ea
and as $|\yv_i| \rightarrow 0$ (see \cite{GR})
\ba\label{intab}
\int \!\!\diff \kv_i \, \f{e^{i \yv_i \cdot \kv_i} (\kv_i \cdot \av +b)}{\kv_i^2 ( \kv_i^2 + \kv_i \cdot \av +b)}&=& \int \!\!\diff \kv_i \, \f{ \kv_i \cdot \av +b}{\kv_i^2 ( \kv_i^2 + \kv_i \cdot \av +b)}	\nonumber	\\
 &+& \int \!\!\diff \kv_i \, \f{(e^{i \yv_i \cdot \kv_i} -1) (\kv_i \cdot \av +b)}{\kv_i^2 ( \kv_i^2 + \kv_i \cdot \av +b)}\nonumber\\
&=& \pi^2 \sqrt{4b - \av^2} + o(1)\nonumber %\\
%&=& 2\pi^2  \f{\sqrt{1+2 \alpha}}{1+\alpha} \sqrt{\breve{\kv}_i^2 +\f{2\alpha}{1+2\alpha} \sum_{j<l;j,l\neq i} \kv_j \cdot \kv_l } \;+\;o(1)
\ea
Using (\ref{intg}) and (\ref{intab}) in (\ref{A}) we obtain (see (\ref{elle}))
\newpage
\bml{
A= \sqrt{\f{2}{\pi}} \f{(-1)^{i+1}}{(2 \pi)^{\f{3}{2}N}} \disp\int\!\!\diff\breve{\kv}_i \, e^{i \breve{\yv}_i \cdot \breve{\kv}_i} \hat{\xi} (\breve{\kv}_i) \times	\\
\times \left( \f{2 \pi^2}{|\yv_i|}- 2 \pi^2 \f{\sqrt{1 + 2 \alpha}}{1 +\alpha} L(\breve{\kv}_i) + o(1) 
\right)	\\
= \f{(-1)^{i+1} \xi(\breve{\yv}_i)}{|\yv_i|}	- \f{\sqrt{1 + 2 \alpha}}{1+\alpha} \f{(-1)^{i+1}}{(2 \pi)^{\f{3}{2}(N-1)}}  \times	\\
\times \int\!\!\diff\breve{\kv}_i \, e^{i \breve{\yv}_i \cdot \breve{\kv}_i} \hat{\xi} (\breve{\kv}_i) \, L(\breve{\kv}_i)+ o(1).
}
Considering  the contribution of the other terms of the sum in (\ref{poten}), it is now easy to derive the asymptotic behavior (\ref{asy}). 

Let us verify that the potential ${\cal G} \xi$ satisfies  equation \eqref{distri}. From the definition \eqref{poten} and the expression of $H_0$ in Fourier transform we have 
\ba
&&(H_0 \mathcal G \xi)(\yyv)= \f{1}{(2 \pi)^{\f{3}{2}N} } \sqrt{\f{2}{\pi}} \sum_{j=1}^N  (-1)^{j+1} \!\!\!\int_{\R^{3N}} \!\!\! \diff \bf{K}\,     e^{i \bf{K} \cdot \yyv} \hat{\xi}(\breve{\kv}_j) \nonumber\\
&&=\f{1}{(2 \pi)^{\f{3}{2}} } \sqrt{\f{2}{\pi}} \sum_{j=1}^N (-1)^{j+1} \xi(\breve{\yv}_j) \int_{\R^{3}} \!\!\! \diff \kv_j\, e^{i \kv_j \cdot \yv_j}  \nonumber\\
&&= 4 \pi \sum_i (-1)^{i+1}\xi(\breve{\yv}_i) \, \delta(\yv_i)
\ea

%\n

\vs
\begin{acknowledgements}
M.C. and D.F.  acknowledge the support of MIUR through the FIR grant 2013 ``Condensed Matter in Mathematical Physics (Cond-Math)'' (code RBFR13WAET) and the FIRB grant 2012 ``Dispersive dynamics: Fourier analysis and variational methods''.  
\end{acknowledgements}

\vs\vs

\end{document}